\documentclass[aps,prd,twocolumn,superscriptaddress,longbibliography,nofootinbib]{revtex4-2}
\usepackage[utf8]{inputenc}
\usepackage{color}
\usepackage{graphicx}
\usepackage{subfigure}
\usepackage{xcolor}
\usepackage[normalem]{ulem}
\usepackage[pdftex,breaklinks,colorlinks,linkcolor=blue,citecolor=teal,anchorcolor=red,urlcolor=cyan]{hyperref}
\usepackage{orcidlink}
\usepackage{multirow}

\usepackage{amsmath,amssymb,amsfonts}

\def\prl{Phys. Rev. Lett.}
\def\prd{Phys. Rev. D}

\def\apj{Astrophys. J.}

\def\apjl{Astrophys. J. Lett.}

\def\pau_p{Prog. Theor. Phys.}

\def\mnras{Mon. Not. R. Astron. Soc.}

\def\apss{Astrophys. Space Sci.}
\def\physrep{Phys. Rep.}
\def\nat{Nature}
\def\sovast{Soviet Ast.}
\def\jcap{J. Cosmology Astropart. Phys}

\def\eq{Eq.}
\def\P{\mathcal P}

\begin{document}

\title{Primordial black holes captured by neutron stars: simulations in general relativity}

\author{Thomas W.~Baumgarte\orcidlink{0000-0002-6316-602X}}
\email{tbaumgar@bowdoin.edu}
\affiliation{Department of Physics and Astronomy, Bowdoin College, Brunswick, Maine 04011, USA}

\author{Stuart L.~Shapiro\orcidlink{0000-0002-3263-7386}}
\email{slshapir@illinois.edu}
\affiliation{Department of Physics, University of Illinois at Urbana-Champaign, Urbana, Illinois 61801}
\affiliation{Department of Astronomy and NCSA, University of Illinois at Urbana-Champaign, Urbana, Illinois 61801}

\begin{abstract}
We present self-consistent numerical simulations in general relativity of putative primordial black holes inside neutron stars.  Complementing a companion paper in which we assumed the black hole mass $m$ to be much smaller than the mass $M_*$ of the neutron star, thereby justifying a point-mass treatment, we here consider black holes with masses large enough so that their effect on the neutron star cannot be neglected.  We develop and employ several new numerical techniques, including initial data describing boosted black holes in neutron-star spacetimes, a relativistic determination of the escape speed, and a gauge condition that keeps the black hole hole at a fixed coordinate location.  We then perform numerical simulations that highlight different aspects of the capture of primordial black holes by neutron stars.  In particular, we simulate the initial passage of the black hole through the star, demonstrating that the neutron star remains dynamically stable provided the black-hole mass is sufficiently small,  $m \lesssim 0.05 M_*$. We also model the late evolution of a black hole oscillating about the center of an initially stable neutron star while accreting stellar mass and ultimately triggering gravitational collapse.
\end{abstract}

\maketitle

%
\section{Introduction}
%

First proposed by \cite{ZelN67,Haw71,CarH74}, {\em primordial black holes} (PBHs) may or may not have formed in the early Universe.  While there is no direct evidence for their formation (but see \cite{CarCGHK24} for possible indirect indications), they may account for the Universe's dark matter content either in its entirety or in part.  A number of different considerations and observations limit the possible contribution of PBHs to the dark matter for different PBH masses, but other mass ranges remain unconstrained (see, e.g., \cite{Khl10,CarKSY21,CarK20} for reviews; see also \cite{MonCFVSH19}).   Fig.~10 in \cite{CarKSY21} and Fig.~1 in \cite{CarK20}, for example, identify several mass windows in which PBHs remain viable dark-matter candidates.

Numerous authors have explored possible consequences and observational signatures of PBHs, ranging from interactions with the Earth (e.g.~\cite{JakR73,BeaT74,LuoHTP21}) and  solar-system \cite{BerCDVC23,TraGLK23} to a host of high-energy astrophysical scenarios, including neutron star implosions \cite{FulKT17,BraLT18}, fast radio bursts \cite{FulKT17,AbrBW18,AmaS23}, long-period transients \cite{BauS24a}, the formation of low-mass stellar black holes \cite{Tak18,TakFK21,AbrBUW22,OncMGG22}, microquasars \cite{Tak19}, and the collapse to supermassive black holes (e.g.~\cite{BamSDFV09}), possibly via the formation of PBH clusters \cite{BelDEGGKKRS14,BelDEEKKKNRS19}. Many of these scenarios invoke the {\em capture} of a PBH by a star, for example following their collision.  Specifically, a PBH that traverses a star loses energy due to dissipative forces, most importantly dynamical friction and a drag force resulting from accretion.  It turns out that these dissipative forces are most efficient in {\em neutron stars}, which are therefore the most likely candidates to bind a PBH gravitationally (see, e.g.,~\cite{CapPT13,AbrBW18,GenST20,AbrBUW22,CaiBK24,BauS24c}).   The PBH may still reemerge after the first passage, but, if gravitationally bound, will return to the neutron star for subsequent passages, and will at some point have lost enough energy so that it can no longer emerge and is instead confined to the stellar interior.  Once confined, the PBH will settle down to the host star's center.  During its inspiral it will emit a continuous gravitational wave signal which, if detected, would reveal properties of the neutron star interior and the nuclear equation of state (e.g.~\cite{HorR19,ZouH22,GaoDGZZZ23,BauS24b}).  The PBH also continues to accrete stellar material, ultimately triggering dynamical gravitational collapse and terminating the co-evolution of the host star with an ``endo-parasitic" black hole (see also \cite{EasL19,RicBS21b,SchBS21} for numerical simulations).  

While collisions between PBHs and neutron stars are expected to be rare (e.g.~\cite{Abretal09,CapPT13,HorR19,ZouH22,CarCGHK24}), estimates of these rates depend strongly on a number of assumptions and may be more favorable in special environments, e.g.~globular clusters and galactic centers.   While we focus on direct collisions here, PBHs may also be captured by neutron stars by means of other processes (e.g.~\cite{BamSDFV09,PanL14,HorR19,GenST20}).  

That capture and confinement of low-mass PBHs by neutron stars, together with the PBH's subsequent accretion and the neutron star's dynamical collapse, is governed by processes that act on vastly differing length- and time-scales.  Accordingly, different aspects of this problem have to be treated using different approaches and approximations (see, e.g., Section II in \cite{BauS24c}, hereafter Paper I, for a detailed discussion).  While the mass $m$ of the black hole is small compared to the mass $M_*$ of the neutron star, $m \ll M_*$, effects of the PBH on the neutron star can be neglected in a first approximation. In this case, the PBH can be modeled as a point-mass in a fixed neutron-star background, allowing for long-time simulations of the capture, confinement, and accretion processes.  This is the approach that we adopted in Paper I.

Complementing the treatment of Paper I, we here consider larger PBH masses with $m \simeq 0.01 M_*$. In this case the effects of the black hole on the neutron cannot be neglected, so that fully self-consistent numerical simulations are required to model these interactions.  We perform such simulations both for an initial collision of a black hole with a neutron star, and for the late evolution prior to dynamical collapse.  Regarding the former, we explore the mass limit above which a black hole would induce dynamical collapse during its first transit through a neutron star.  In particular, we demonstrate that a neutron star will remain dynamically stable despite having been pierced by a black hole, provided the black-hole mass is sufficiently small.\footnote{Compare with \cite{CarIZZ22,ZhoCIZ23}, who considered boson stars pierced by black holes.}  Regarding the latter we generalize the simulations of \cite{EasL19,RicBS21b,SchBS21}, which assumed the black hole to reside at the neutron-star center, and
consider a black hole orbiting about the center while accreting stellar material and ultimately triggering gravitational collapse.  

Our paper is organized as follows.  In Section \ref{sec:num} we introduce several numerical techniques adopted in our simulations.  In particular, we discuss the construction of initial data describing boosted black holes in neutron-star spacetimes, including a self-consistent, relativistic determination of the escape speed (Section \ref{sec:num:indata}), the evolution calculation implementing a ``fix-point" gauge condition that keeps the black hole at the origin of our spherical polar coordinate system (Section \ref{sec:num:evolve}), as well as several useful diagnostics (Section \ref{sec:num:diagnostics}).  In Section \ref{sec:results} we present numerical results both for the simulations of the initial transit (Section \ref{sec:res:transit}) as well as the late accretion (Section \ref{sec:res:late}).  We briefly summarize our results in Section \ref{sec:discussion}.  Finally, in Appendix \ref{sec:accretion} we discuss the difference between rest-mass and mass-energy accretion, illustrated by spherically symmetric, stationary Bondi flow.  Unless stated otherwise we adopt geometrized units with $G = c = 1$.

%
\section{Numerical Setup}
\label{sec:num}
%

%
\subsection{Initial data}
\label{sec:num:indata}
%

%
\subsubsection{Solving the constraint equations}
\label{sec:num:constraints}
%

We set up initial data by generalizing the approach of \cite{RicBS21b}, who assumed the black hole to be at rest with respect to the host star and located at its center.  Here we generalize both of these assumptions.

We adopt a frame in which the star is at rest, so that its momentum density vanishes, $S^i = 0$, and place the black hole at the star's surface in one set of simulations and well inside the star in another.  We assign the black hole a 3-momentum $\P^i$ directed along the radius of the star, so that simulations that track the evolution of the system can be performed in axisymmetry. The scenario we consider thus applies to head-on collisions.  We will also assume maximal slicing, so that the mean curvature vanishes, $K = 0$, as well as conformal flatness so that we may write the spatial metric as
\begin{equation}
\gamma_{ij} = \psi^4 \eta_{ij},
\end{equation}
where $\psi$ is the conformal factor and $\eta_{ij}$ the flat metric.   Analytical solutions to the momentum constraints describing a black hole with a momentum $\P^i$ is then given by the Bowen-York solution
\begin{equation}  \label{BY1}
\tilde A^{ij} = \frac{3}{2r^2} \left(\P^i l^j + \P^j l^i + (l^i l^j - \eta^{ij}) \, l_k \P^k \right)
\end{equation}
(see \cite{BowY80}), where $r$ measures the coordinate distance from the black-hole puncture at the origin of the coordinate system,  $l^i$ is the unit vector pointing away from the puncture (i.e.~$l^i = x^i/r$ in Cartesian coordinates, but $l^i = (1,0,0)$ in spherical polar coordinates centered on the puncture), and $\tilde A^{ij}$ is the conformally rescaled, trace-free part of the extrinsic curvature, so that
\begin{equation}
K_{ij} = \psi^{-2} \tilde A_{ij}
\end{equation}
(since $K = 0$).  

Under these assumptions the Hamiltonian constraint takes the form
\begin{equation} \label{Ham1}
\tilde \nabla^2 \psi + \frac{1}{8} \psi^{-7} \tilde A_{ij} \tilde A^{ij} = - 2 \pi \psi^{5 + k} \tilde \rho,
\end{equation}
where $\tilde \nabla^2$ is the flat Laplace operator and where we have rescaled the total mass-energy density as observed by a normal observer, $\rho$, according to 
\begin{equation} \label{rho_conformal_rescaling}
\rho = \psi^k \tilde \rho.
\end{equation}
In practice we choose $k = -6$ in the following.  We also note that \begin{equation}
\tilde A_{ij} \tilde A^{ij} = \frac{9 \P^2}{2 r^4} (1 + 2 \cos^2 \theta)
\end{equation}
for the Bowen-York solution (\ref{BY1}), where $\theta$ is the angle between $l^i$ and $\P^i$ and $\P^2 = \eta_{ij} \P^i \P^j$. 

\begin{table}[t]
    \centering
    \begin{tabular}{c|c|c|c|c|c|c|c}
         Model & $\bar \rho_{0c}$ &  $\bar M_*$  & $\bar M_0$  &  $\bar R_*$    &  $\bar r_*$ & $|\P|/m$ & $t_0$~[ms]\\
         \hline
       A & 0.126 & 0.139 & 0.15 & 0.962 & 0.817 & 0.74 & 0.127\\
        B & 0.152 & 0.148 & 0.16 & 0.925 & 0.772 & 0.80 & 0.110\\
        C & 0.190 & 0.156 & 0.17 & 0.889 & 0.716 & 0.875 & 0.096\\
    \end{tabular}
    \caption{Properties of selected OV solutions for $n = 1$, $\Gamma = 2$ polytropes close to the maximum allowed rest mass $\bar M_0 = 0.1799$ and total mass-energy $\bar M = 0.1637$.  For each model we list the central rest-mass density $\bar \rho_{0c}$, the gravitational (ADM) mass $\bar M_*$, the rest mass $\bar M_0$, the areal radius $\bar R_*$, the isotropic radius $\bar r_*$, as well as escape speed $|\P|/m$ from the stellar surface as determined in Section \ref{sec:num:find_momentum}.  In the last column we list the dynamical timescale $t_0$ for stars with mass $M_* = 1.4 M_\odot$ (see Eq.~\ref{t0_cgs}).}
    \label{tab:TOV_solutions}
\end{table}

We start with a solution to the Oppenheimer-Volkoff (OV) equations \cite{OppV39}, describing a spherical equilibrium star centered on some location $z_{\rm TOV}$ on the $z$-axis for a given central rest-mass density $\rho_{0c}$.   In this paper we assume that, initially, the star is governed by a polytropic equation of state (EOS)
\begin{equation} \label{eos}
P = \kappa \rho_0^{\Gamma},~~~~~~~~\Gamma = 1 + 1/n,
\end{equation}
where $P$ is the pressure, $\kappa$ is a (dimensional) constant, and $n$ the polytropic index.  Since the constant $\kappa$ must have units of length$^{2/n}$ in geometrized units, we may define dimensionless quantities by scaling out suitable powers of $\kappa^{n/2}$ (see, e.g., Section 1.3 in \cite{BauS10}).  For example, we define
\begin{equation} \label{scaling}
\bar R = \kappa^{- n/2} R,~~~~ \bar M = \kappa^{-n/2} M,~~~~~
\bar \rho = \kappa^n \rho.
\end{equation}
In order to model a moderately stiff EOS we adopt $n = 1$ corresponding to $\Gamma = 2$.  In Table \ref{tab:TOV_solutions} we list properties of three specific OV solutions, labeled A, B, and C, close to the maximum allowed rest mass for this EOS.

The OV solution then provides both an energy density 
\begin{equation} \label{rho_NS}
\tilde \rho_{\rm NS} = \psi^{-k}_{\rm NS} \rho_{\rm NS}
\end{equation}
and a conformal factor $\psi_{\rm NS}$ that satisfy (\ref{Ham1}) for vanishing $\bar A_{ij}$, i.e.
\begin{equation} \label{Ham_NS}
\tilde \nabla^2 \psi_{\rm NS} = - 2 \pi \psi^{5 + k} \tilde\rho_{\rm NS}.
\end{equation}
A static black hole in the absence of a star, on the other hand, satisfies the Hamiltonian constraint (\ref{Ham1}) for
\begin{equation} \label{psi_BH}
\psi_{\rm BH} = 1 + \frac{{\mathcal M}}{2r},
\end{equation}
where ${\mathcal M}$ is the black hole's puncture mass.  

In order to construct a solution that describes a boosted black hole inside a star we now generalize the puncture method of \cite{BraB97} and write the conformal factor $\psi$ as
\begin{equation} \label{psi}
\psi = \psi_{\rm NS} + \psi_{\rm BH} + u - 1,
\end{equation}
where $u$ is a (regular) correction that accounts for both the black hole's boost and the nonlinear interaction between the black hole and star.  Note also that the ansatz (\ref{psi}) allows $u \rightarrow 0$ asymptotically, assuming that $\psi$, $\psi_{\rm NS}$, and $\psi_{\rm BH}$ all approach unity there.  Inserting the ansatz (\ref{psi}) into the Hamiltonian constraint (\ref{Ham1}) we then obtain an equation for the correction $u$,
\begin{align} \label{Ham_u}
\tilde \nabla^2 u = & - 2 \pi \left((\psi_{\rm NS} + \psi_{\rm BH} + u - 1)^{5+k} - \psi_{\rm NS}^{5+k} \right) \tilde \rho_{\rm NS} \nonumber \\
& - \frac{1}{8} (\psi_{\rm NS} + \psi_{\rm BH} + u - 1)^{-7} \tilde A_{ij} \tilde A^{ij},
\end{align}
where we have used (\ref{Ham_NS}) together with $\tilde \nabla^2 \psi_{\rm BH} = 0$ and where we assume that $\tilde \rho = \tilde \rho_{\rm NS}$ is being held fixed.  We solve (\ref{Ham_u}) to a desired tolerance, and then construct the conformal factor $\psi$ from (\ref{psi}).  Given $\psi$, we compute the new (physical) energy density 
from Eqs.~(\ref{rho_conformal_rescaling}) and (\ref{rho_NS},
\begin{equation}
\rho = \psi^k \tilde \rho = \left( \frac{\psi}{\psi_{\rm NS}} \right)^k \rho_{\rm NS}.
\end{equation}
Note, in particular, that $\rho$ vanishes at the black-hole puncture, where $\psi \rightarrow \infty$, for $k < 0$.  We also note that, for a given central density $\rho_{0c}$, the rest mass of the neutron star in the presence of a black hole will be different from that listed in Table~\ref{tab:TOV_solutions}, i.e.~in the absence of the a black hole. 

We also determine the initial lapse function $\alpha$ from the ``pre-collapsed" lapse condition $\alpha = \psi^{-2}$, and set the initial shift vector $\beta^i$ to zero.

Examples of initial data constructed as described above are shown in the top left panels of Figs.~\ref{fig:Run_B1}, \ref{fig:Run_B2}, and~\ref{fig:Run_B3}.

%
\subsubsection{Determining the black-hole momentum}
\label{sec:num:find_momentum}
%

\begin{figure}
\centering
\includegraphics[width = 0.45 \textwidth]{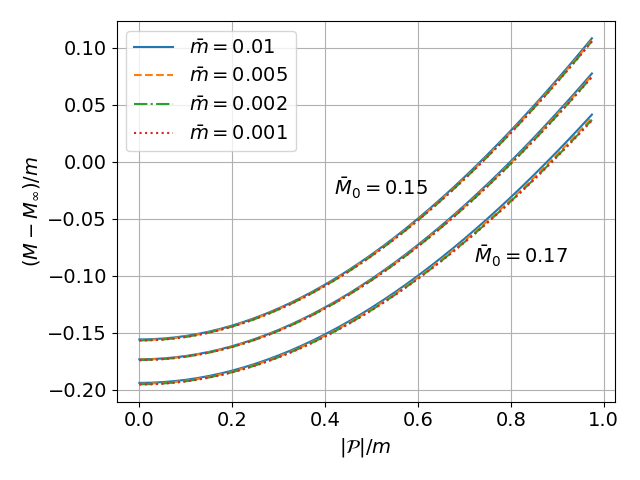}
\caption{The difference between the total ADM mass $M$ and its value at infinite separation, $M_{\infty}$, divided by the black hole mass $m$, as a function of the magnitude of the black-hole momentum $|\P|$.  We show results for different black hole masses $\bar m$ and the three neutron star models listed in Table \ref{tab:TOV_solutions} with rest masses $\bar M_0 = 0.15$, 0.16, and 0.17.  For a given value of $\bar M_0$, the lines for different $\bar m$ can hardly be distinguished.  Locating the zeros of $M - M_\infty$ we identify the escape speeds $|\P|/m$ for each $\bar M_0$.  As in the geodesic point-mass limit these values are nearly independent of $\bar m$ for the examples considered here.}
\label{fig:find_v0}
\end{figure}

In order to construct initial data that faithfully represent a black hole entering the neutron star following free-fall from a large separation (see Section \ref{sec:res:transit}) we need to make realistic choices for the black-hole's initial momentum $\P^z$.  

In a Newtonian context this momentum can be estimated using a point-mass approximation together with conservation of energy.  Assuming that the relative speed between the black hole and the neutron star at large separation is small in comparison to the speed at which the black hole will reach the stellar surface, the latter is given by the escape speed  
\begin{equation} \label{escape_speed_Newton}
v_{\rm esc} = \frac{\P}{m} = \left(\frac{2 M_*}{r_*} \right)^{1/2}~~~~~\mbox{(Newtonian),}
\end{equation}
where we have assumed that the black hole mass $m$ is much smaller than that of the neutron star, $m \ll M_*$.  In particular, we note that $v_{\rm esc}$ is independent of $m$ in this limit.

In a relativistic context one could adopt a similar point-mass approximation.  Using the fact that the energy per unit mass $e \equiv - u_t$ is conserved for geodesic motion one can obtain an expression for the black-hole's four-velocity at the surface of the neutron star.  This approach still uses a point-mass approximation, however, and moreover it is not clear a priori how to translate the four-velocity of a particle inside a gravitational potential into the momentum $P^i$ in (\ref{BY1}), i.e.~the momentum of the black hole as observed by a distant observer.

We therefore adopt a similar energy conservation argument applied to our sets of initial data; specifically, we will construct initial data that, for given black-hole and neutron-star masses, have the same ADM mass at finite separation as they would have at infinite separation.

As before, we assume that both the neutron star and black hole are at rest when at infinite separation, so that their total ADM energy is given by 
\begin{equation}
M_\infty = M_{*} + m.
\end{equation}
Here $M_*$ is the neutron star's gravitational mass, and $m$ the black hole's irreducible mass
\begin{equation}
m = \left( \frac{\mathcal A}{16 \pi} \right)^{1/2},
\end{equation}
where we estimate the proper horizon area ${\mathcal A}$ from that of the apparent horizon.  At infinite separation, the irreducible mass $m$ is equal to the puncture mass ${\mathcal M}$ that appears in (\ref{psi_BH}), but not at finite separations.

Neglecting the small amount of gravitational radiation energy emitted by the black hole, we assume that its irreducible mass and the neutron star's baryon number remain conserved during the infall. We next place a black hole with the same irreducible mass $m$ close to the surface of a neutron star with the same rest mass $M_0$ (by centering the neutron star on a location $z_{\rm OV} = - r_*$ as listed in Table \ref{tab:TOV_solutions} for the chosen neutron-star mass).  In practice, we choose a value for the momentum $P^z$, and then iterate over both ${\mathcal M}$ in Eq.~(\ref{psi_BH}) and the central rest-mass density $\rho_{0c}$ in the OV solution.  For each guess ${\mathcal M}$ and $\rho_{0c}$ we solve the constraint equations as described in Section \ref{sec:num:constraints}, measure the resulting irreducible mass $m$ and rest mass $M_0$, and continue the iteration until the desired values have been achieved to within a given tolerance.  Finally we compute the ADM mass $M$, which provides the total gravitational energy for the given value of $P^z$. This ADM mass $M$ should be the same as its counterpart for infinite separation $M_\infty$.

In Fig.~\ref{fig:find_v0} we show results for $(M - M_\infty)/m$ as a function of $\P^z$ for different black hole masses $m$ and for the three different stellar models listed in Table \ref{tab:TOV_solutions}.  Locating the zeros of $M - M_\infty$ then identifies the escape speeds $|\P|/m$, which we have included in Table \ref{tab:TOV_solutions}.

%
\subsection{Evolution}
\label{sec:num:evolve}
%

%
\subsubsection{Numerical code}
\label{sec:num:code}
%

We evolve the initial data of Section \ref{sec:num:indata} using a finite-difference code that implements the Baumgarte-Shapiro-Shibata-Nakamura (BSSN) formulation of Einstein's equations \cite{NakOK87,ShiN95,BauS99} in spherical polar coordinates.   The code adopts a reference-metric formulation (e.g., \cite{BonGGN04,ShiUF04,Bro09,Gou12}) together with a rescaling of all tensorial quantities in order to handle the coordinate singularities at the origin $r = 0$ and the axis where $\sin \theta = 0$ (see \cite{BauMCM13,BauMM15}).  We apply similar techniques to the equations of relativistic hydrodynamics (see \cite{MonBM14}) and use a Harten-Lax-van-Leer-Einfeld approximate Riemann solver \cite{HarLL83,Ein88} together with a simple monotonized central-difference limiter reconstruction scheme \cite{Van77} to handle shocks.  During the evolution we adopt a $\Gamma$-law equation of state
\begin{equation} \label{Gamma_law}
P = (\Gamma - 1) \rho_0 \epsilon,
\end{equation}
where $\epsilon$ is the specific internal energy density, and where we again choose $\Gamma = 2$.

For the simulations shown in this paper we use fourth-order finite-differencing for spatial derivatives, together with a fourth-order Runge-Kutta method for the time evolution.  While the code does not assume any symmetries, our simulations are performed in axisymmetry, so that only a single (interior) grid-point is required to resolve the azimuthal angle $\varphi$.  For the polar angle $\theta$ we use a grid with $N_\theta$ grid-points uniformly distributed in $\theta$.  Our radial grid extends from $r = 0$ to $r = r_{\rm out}$ and is constructed from a uniform, cell-centered grid with $N_r$ grid-points in an auxiliary variable $0 \leq x \leq 1$ with the map
\begin{equation}
 r = r_{\rm out} \frac{\sinh (s_r x)}{\sinh (s_r)}
\end{equation}
where $s_r$ is a constant parameter that governs the ``non-uniformity" of the grid  (see \cite{RucEB18}).  For $s_r = 0$ we recover a uniform grid in $r$, while for $s_r > 0$ the grid is nearly uniform close to the origin at $r = 0$, but becomes approximately logarithmic far from the origin.

Because of resource limitations we performed all simulations on quite coarse numerical grids.  Specifically, in Section \ref{sec:res:transit} we used $\bar r_{\rm out} = 20$, $N_r = 192$, $N_\theta = 32$, and we adjusted the parameter $s_r$ so that the interior of the black hole is covered by at least about 12 grid-points.\footnote{The cases C in Section \ref{sec:res:transit}, with the neutron star masses closest to the maximum allowed mass, turned out to be most sensitive to numerical error and sometimes required slightly higher resolution close to the black hole, in which case we reduced $N_\theta$ to 24 in order to speed up the time evolution.}  For the simulations in Section \ref{sec:res:late}, for which the black hole is always close to the center of the neutron star, the angular dependence of all functions is relatively small, so that we were able to use $N_\theta = 16$.

%
\subsubsection{Gauge conditions}
\label{sec:num:gauge}
%

In order to take full advantage of the spherical polar coordinates used in our code, it is desirable to keep the location of the black-hole puncture at the origin of the coordinate system for accuracy. 

At the black hole puncture, the conformal factor $\psi$ diverges, and equivalently the function $\chi \equiv \psi^{-4}$ vanishes.  Since $\chi$ satisfies the evolution equation
\begin{equation}
\partial_t \chi = \frac{2}{3} \chi \,( \alpha K - \partial_i \beta^i) + \beta^i \partial_i \chi
\end{equation}
(see, e.g., \eq~(2) in \cite{CamLMZ06}), we see that $\chi$ will remain zero at the current location of the puncture if the shift vector $\beta^i$ vanishes there.  Therefore, a black-hole puncture will remain at a fixed coordinate location $x_{\rm FP}^i$ if we impose $\beta^i = 0$ at that location, which we will refer to as a {\em fixed point} in the following (compare the ``fixed puncture" method of \cite{BruTJ04}).\footnote{We note that the above argument assumes that $\chi$ is differentiable at the puncture.  For trumpet geometries we have $\psi \propto r^{-1/2}$ (see, e.g., \cite{HanHPBM07,HanHOBO08,BauN07,DenB14}  as well as \cite{BauS10,BauS21b} for textbook examples), in which case $\chi \propto r^2$ is indeed differentiable at $r = 0$.}

In order to impose such a fixed point, while simultaneously taking advantage of the desirable properties of moving-puncture coordinates, we choose the following conditions on the lapse and the shift.  We evolve the lapse $\alpha$ with the {\em 1+log} slicing condition
\begin{equation}
\partial_t \alpha = - 2 \alpha K + \beta^i \partial_i \alpha ,
\end{equation}
(see \cite{BonMSS95}) starting with a ``pre-collapsed" lapse $\alpha = \psi^{-2}$ at the initial time $t = 0$.  We also
adopt a {\em Gamma-driver} condition for the shift vector $\beta^i$ (see \cite{AlcBDKPST03}), starting with $\beta^i = 0$ initially, but modify this condition as follows.  

Whenever the time derivative of the shift is needed, we first evaluate a preliminary time derivative, $\dot \beta^i_{\rm GD}$, according to the Gamma-driver condition as suggested by \cite{ThiBHBR11},
\begin{equation} \label{GD_1}
\partial_t \beta^i_{\rm GD} = \mu_S \tilde \Gamma^i - \eta \beta^i +  \beta^j \partial_j \beta^i,
\end{equation}
where we adopt $\eta = 0$ and $\mu_S = 0.75$ in our simulations.  We then interpolate $\dot \beta^i_{\rm GD}$ to the location $x_{\rm FP}^i$ of the fixed point, resulting in $\dot \beta^i_{\rm GD}(x_{\rm FP}^i)$.  In our axisymmetric simulations, with the black hole and $x_{\rm FP}^i$ on the symmetry axis, we expect $\dot \beta^i_{\rm GD}(x_{\rm FP}^i)$ to be aligned with the axis -- meaning that, in Cartesian coordinates, it would have a non-zero $z$-component $\dot \beta^z_{\rm GD}(x_{\rm FP}^i)$ only.

We now interpret $\dot \beta^i_{\rm GD}(x_{\rm FP}^i)$ as a ``constant" vector field, by which we mean that its Cartesian components are the same everywhere.  In the spherical polar coordinates of our code we use the flat-space transformation laws
\begin{align}
\dot \beta^r_{\rm GD}(x_{\rm FP}^i) & = \cos \theta \, \dot \beta^z_{\rm GD}(x_{\rm FP}^i), \\
\dot \beta^\theta_{\rm GD}(x_{\rm FP}^i) & = - \sin \theta \, \dot \beta^z_{\rm GD}(x_{\rm FP}^i) / r
\end{align}
to compute the components $\dot \beta^i_{\rm GD}(x_{\rm FP}^i)$ from $\dot \beta^z_{\rm GD}(x_{\rm FP}^i)$ everywhere.

In a second step we subtract $\dot \beta^i_{\rm GD}(x_{\rm FP}^i)$ from the values of the time derivatives computed in (\ref{GD_1}) (i.e., we ``shift the shift").  Specifically, we write our final expression for the time derivative of the shift as
\begin{equation} \label{shift_condition}
\partial_t \beta^i = \dot \beta^i_{\rm GD} - f(x^i) \, \dot \beta^i_{\rm GD}(x_{\rm FP}^i),
\end{equation}
where $f(x^i)$ is a yet-to-be determined function of the spatial coordinates.  Choosing $f = 1$ at the fixed point $x_{\rm FP}^i$ means that $\dot \beta^i = 0$ there, and, since $\beta^i = 0$ initially, it will vanish there at all times, as desired.  Moreover, if $f = 1$ everywhere, the actual shift vector used in the code differs from the Gamma-driver shift by a constant vector only, so that it inherits all the desirable properties of the Gamma-driver for black-hole evolutions.  However, our code assumes that $\beta^i \rightarrow 0$ asymptotically, so that this choice would result in an inconsistency at the outer boundaries.  As a compromise we therefore choose
\begin{equation} \label{choice_for_f}
f(x^i) = \exp\left(-(r_{\rm FP} / \sigma)^2\right),
\end{equation}
where $r_{\rm FP}$ is the coordinate distance from the fixed-point, and $\sigma$ a constant.  In practice we choose $\sigma = 2$ in our code units, which is greater than the stellar radius (so that star and black hole are advected more or less uniformly), but smaller than the distance to the outer boundary (so that $f \simeq 0$ there).

\begin{figure}
\centering
\includegraphics[width = 0.45 \textwidth]{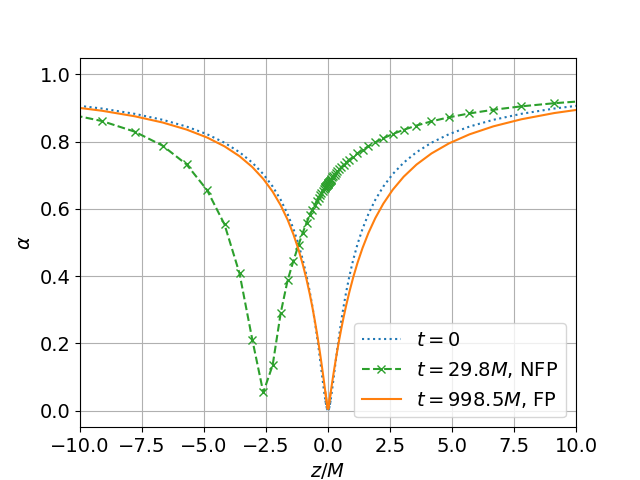}
\caption{The lapse function $\alpha$ along the symmetry axis in an evolution of a black hole with momentum $\P^z = - 0.1 M$ in vacuum with and without fix point.  The initial data, shown as the dotted line, are identical for both evolutions.  In the evolution without fix point (NFP, dashed line) the black hole moves away from the origin, and rather quickly loses resolution.  The evolution crashes not long after the time at which we show the data.  In the evolution with a fix point (FP, solid line), on the other hand, the black hole remains at the origin.  We visualize the radial grid in our evolution by marking individual grid points for the evolution without fix point.}
\label{fig:fixpointtest}
\end{figure}

As a test and illustration of this approach we show in Fig.~\ref{fig:fixpointtest} the lapse function $\alpha$ in the evolution of a black hole carrying momentum $P^z = -0.1 M$ both with and without a fix point.  In the evolution without a fix point, the black hole moves away from the origin, as expected, and soon loses sufficient numerical resolution.  The evolution crashes at a time soon after that shown in the figure.  In the evolution with the fix point, on the other hand, the black hole remains at the origin, and remains well resolved throughout the evolution.

%
\subsection{Diagnostics}
\label{sec:num:diagnostics}
%

%
\subsubsection{Black-hole accretion rate}
\label{sec:num:accretion}
%

We identify the world-tube ${\mathcal F}$ formed by the black-hole apparent horizon ${\mathcal H}$ with the level surface $f = 0$ of the function
\begin{equation} \label{f_def}
f(t,r,\theta,\varphi) \equiv
r - h(t,\theta,\varphi) = 0,
\end{equation}
where $h(t,\theta,\varphi)$ measures the apparent horizon's coordinate distance from a center ${\mathcal C}^i$ in the $(\theta,\varphi)$ direction at time $t$.  In the following we will assume that ${\mathcal C}^i$ is the origin of coordinate system.  As derived in Appendix A in \cite{FarLS10}, the accretion rate can then be written as 
\begin{equation} \label{M_dot_1}
\dot m_0 = - \oint_{\mathcal H} \alpha \sqrt{\gamma} \rho_0 u^\mu (\partial_\mu f) J d\theta d\varphi,
\end{equation}
where $J$ is the Jacobian of the transformation from code coordinates $x^i$ to coordinates $(f,\theta,\varphi)$ (see their \eq~(A11)).  Since we use spherical polar coordinates $(r,\theta,\varphi)$ in our code, we have $J = \partial f/\partial r = 1$. 

We also note a subtlety already: the right-hand side of (\ref{M_dot_1}) measures the rate at which rest mass enters the black hole, which is not the same as the rate at which the black hole's irreducible mass $m$ increases (compare Appendix~\ref{sec:accretion}; see also \cite{AguST21,AguTSL21} for a discussion and \cite{RicBS21b} for numerical examples in the context of Bondi accretion).  In order to distinguish these two rates we introduced the symbol $\dot m_0$ to denote the rest-mass accretion rate, even though this should {\em not} be interpreted as the time derivative of the black hole's rest mass $m_0$, since the latter is not defined.

We next assume that the (coordinate) location of the black-hole horizon changes relatively slowly, so that contributions resulting from the time derivative $\partial_t f$ in (\ref{M_dot_1}) can be neglected and the accretion rate reduces to
\begin{equation} \label{M_dot_2}
\dot m_0 = - \oint_{\mathcal H} \alpha \sqrt{\gamma} \rho_0 u^i (\partial_i f) d\theta d\varphi.
\end{equation}

While (\ref{M_dot_2}) could be evaluated as is, it is instructive to rewrite this expression as follows.  For a given (coordinate) time $t$, the spatial slice $\Sigma_t$ can be foliated, at least locally, by the function $f$, resulting in a 2+1 decomposition of the three-dimensional slice $\Sigma_t$.  The (spatial) unit normal $s_i$ on slices of constant $f$ is then given by
\begin{equation}
s_i = \lambda \,\partial_i f,
\end{equation}
where the normalization factor
\begin{equation}
\lambda = \left( \gamma^{ij} (\partial_i f)(\partial_j f)\right)^{-1/2}
\end{equation}
plays the same role in this 2+1 decomposition as the lapse function $\alpha$ does in a 3+1 decomposition of spacetime.  Following this analogy, we may express the determinant of the spatial metric $\gamma$ as
\begin{equation} \label{det}
\sqrt{\gamma} = \lambda \sqrt{{}^{(2)}\gamma},
\end{equation}
where ${}^{(2)}\gamma$ is the determinant of the two-dimensional metric induced on surfaces of constant $f$.  Inserting (\ref{det}) into (\ref{M_dot_2}) we obtain
\begin{equation} \label{M_dot_3}
\dot m_0 = - \oint_{\mathcal H} \alpha \rho_0 u^i d S_i,
\end{equation}
where $d S_i \equiv s_i \sqrt{{}^{(2)}\gamma} \, d\theta d\varphi$ is the outward-oriented surface element on ${\mathcal H}$.  We may therefore interpret the accretion rate as the proper integral of the matter current density's normal component $\rho_0 u^i s_i$ over the horizon ${\mathcal H}$, as one would expect.  For steady-state spherical Bondi accretion onto a Schwarzschild black hole in general relativity, for example, Eq.~(\ref{M_dot_3}) reduces to Eq.~(\ref{M0_dot}) in Appendix \ref{sec:accretion}, derived by a familiar route.

%
\subsubsection{Center of mass}
\label{sec:num:com}
%

\begin{table*}[]
    \centering
    \begin{tabular}{c|c|c|c|c|c|c|c|c}
         Name & $\bar M_0$ & $v_0 = |\P|/m$ & $\bar m$ & $\bar M$ & $m / M_0$ & $\bar \rho_{0c}$ & $\bar {\mathcal M}$ & outcome \\
         \hline \hline
         A1 & \multirow{4}{*}{0.15} & \multirow{4}{*}{0.74} & $1.0 \times 10^{-2}$ & 0.149
         & 0.0667 & 0.1257  & $8.746 \times 10^{-3}$ & collapse during first passage \\
         A2 & &  & $5.0 \times 10^{-3}$ & 0.144 & 0.0333 & 0.1260 & $4.400 \times 10^{-3}$ & collapse after first passage \\
         A3 & &  & $2.0 \times 10^{-3}$ & 0.141 & 0.0133 & 0.1261 & $1.760 \times 10^{-3}$& black hole reemerges without triggering collapse \\
         A4 & &  & $1.0 \times 10^{-3}$ & 0.140 & 0.0067 & 0.1261 & $8.798 \times 10^{-4}$& black hole reemerges without triggering collapse \\
         \hline
         B1 & \multirow{4}{*}{0.16} & \multirow{4}{*}{0.80} & $1.0 \times 10^{-2}$ & 0.158
         & 0.0625 & 0.1504 & $8.682 \times 10^{-3}$ & collapse during first passage \\
         B2 & & & $5.0 \times 10^{-3}$ & 0.153 & 0.0313 & 0.1509 & $4.340 \times 10^{-3}$   & collapse after first passage\\
         B3 & & & $2.0 \times 10^{-3}$ & 0.150 & 0.0125 & 0.1511 & $1.737 \times 10^{-3}$ & black hole reemerges without triggering collapse\\
         B4 & & & $1.0 \times 10^{-3}$ & 0.149 & 0.0063 & 0.1512 & $8.677 \times 10^{-4}$ & black hole reemerges without triggering collapse \\
         \hline
         C1 & \multirow{4}{*}{0.17} & \multirow{4}{*}{0.875} & $1.0 \times 10^{-2}$ & 0.166 
         & 0.0588 & 0.1871  & $8.522 \times 10^{-3}$ & collapse during first passage \\
         C2 & &  & $5.0 \times 10^{-3}$ & 0.161 & 0.0294 & 0.1880 & $4.260 \times 10^{-3}$ & collapse after first passage \\
         C3 & &  & $2.0 \times 10^{-3}$ & 0.158 & 0.0118 & 0.1886 & $1.704 \times 10^{-3}$&  black hole reemerges without triggering collapse\\
         C4 & &  & $1.0 \times 10^{-3}$ & 0.157 & 0.0059 & 0.1888 & $8.517 \times 10^{-4}$&  black hole reemerges without triggering collapse\\
    \end{tabular}
    \caption{Summary of initial data configurations for initial-transit simulations of Section \ref{sec:res:transit}.  For each of the neutron star models $A$, $B$, and $C$ listed in Table~\ref{tab:TOV_solutions}, with rest mass $\bar M_0$ and escape speed $v_0$, we construct initial data for four different irreducible black hole masses $\bar m$, resulting in spacetimes with ADM mass $\bar M$.  Constructing these data entails an iteration over both the rest-mass density at the center of the neutron star, $\bar \rho_{0c}$, and the black-hole puncture mass $\bar {\mathcal M}$.  In the last column we describe the dynamical outcome of the collision.}
    \label{tab:transit}
\end{table*}

We would also like to track the black hole's trajectory through the neutron star.  Given our assumption of axisymmetry the black hole will always remain on the symmetry axis during the head-on collision, so that it is sufficient to determine its $z$-location.  However, using the ``fix-point" shift condition of Section \ref{sec:num:gauge}, the black hole also remains at the origin with $z_{\rm BH} = 0$, so that this by itself is not a useful diagnostic.  Instead, we locate the system's center of mass $z_{\rm CM}$, at least approximately, and then compute the black hole's position relative to the center of mass from
\begin{equation}
z_{\rm BH}^{\rm CM} \equiv z_{\rm BH} - z_{\rm CM}
= - z_{\rm CM}.
\end{equation}

An accurate and invariant determination of the center of mass would entail an expansion of the gravitational fields in the asymptotic region.  For our purely diagnostic purposes here it is sufficient to invoke a much simpler Newtonian and coordinate-based approach.  Specifically, we simply compute
\begin{equation} \label{com}
z_{\rm CM} = \frac{1}{M} \int z \rho_0 d^3x 
= \frac{1}{M_{\rm NS} + m} \int_{\rm NS} z \rho_0 d^3 x,
\end{equation}
where the black hole's contribution to the integrand vanishes because $z_{\rm BH} = 0$, and where we compute $M_{\rm NS} = \int_{\rm NS} \rho_0 d^3x$.

%
\section{Results}
\label{sec:results}
%

For all results shown in this section we construct initial data for an $n = 1$ polytropic EOS (\ref{eos}) as described in Section \ref{sec:num:constraints} and consider the three neutron-star models listed in Table \ref{tab:TOV_solutions}.  In terms of the non-dimensional units introduced in Eq.~(\ref{scaling}), the dynamical timescale $\bar t_0$ for these models can be estimated from
\begin{equation} \label{t0}
\bar t_0 = \left( \frac{\bar R_*^3}{\bar M_*} \right)^{1/2}.
\end{equation}
In order to obtain this timescale in terms of cgs units we write
\begin{equation} \label{t0_cgs}
t_0 = \kappa^{n/2} \, \bar t_0 = \frac{G M_*}{c^3} \left( \frac{\bar R_*}{\bar M_*} \right)^{3/2},
\end{equation}
where we have introduced $G$ and $c$ in the last equality, and where $M_*$ is the stellar mass in terms of cgs units.  In the last column of Table \ref{tab:TOV_solutions} we include values of $t_0$ assuming $M_* = 1.4 M_\odot$.  

In the following we describe two different types of simulations, namely those of the initial transit of a black hole through a neutron star (Section \ref{sec:res:transit}), and an oscillation about the neutron star center mimicking the late-time evolution of a captured black hole (Section \ref{sec:res:late}).

%
\subsection{Initial transit}
\label{sec:res:transit}
%

\begin{figure*}
    \centering
    \includegraphics[width = 0.48 \textwidth]{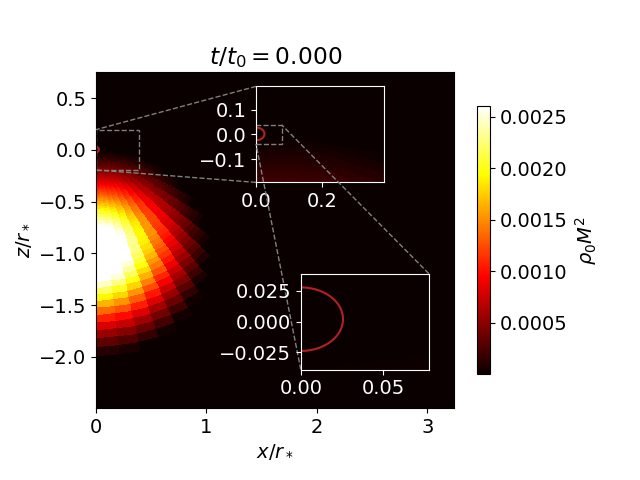}
    \includegraphics[width = 0.48 \textwidth]{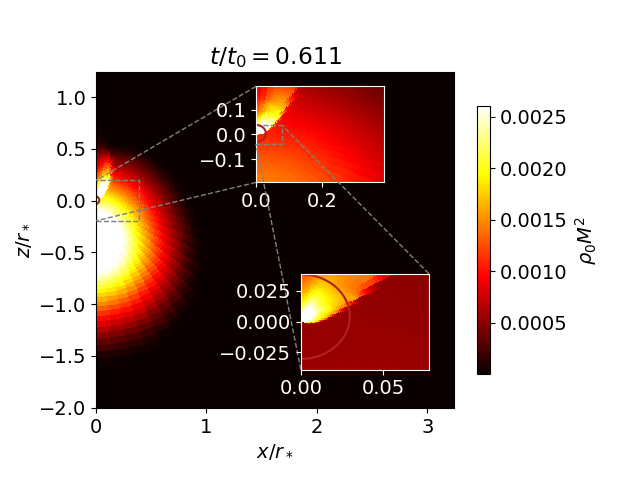}

    \includegraphics[width = 0.48 \textwidth]{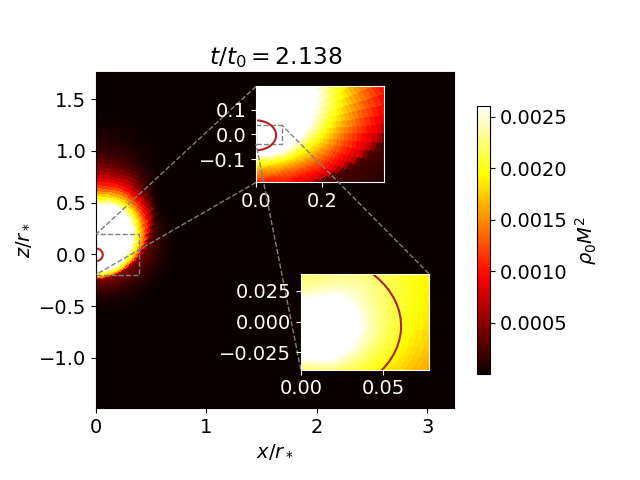}
    \includegraphics[width = 0.48 \textwidth]{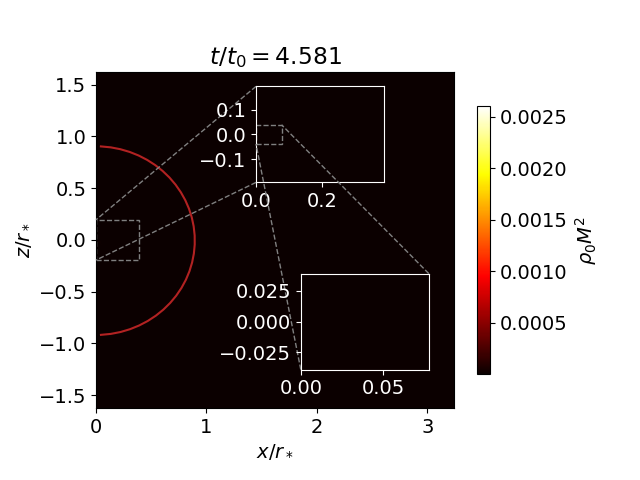}
    \caption{Snapshots of density profiles for a collision of an $\bar m = 10^{-2}$ black hole with a neutron star of mass $\bar M_0 = 0.16$ (Case B1 in Table \ref{tab:transit}).  The color bar refers to densities in the main plots; in order to obtain better contrasts in the insets we allowed for higher densities there.  The red line marks the apparent horizon, and time is recorded in units of the dynamical timescale $t_0$ as defined in (\ref{t0}).  The initial data at $t = 0$ (top left) describe the black hole at the stellar surface with an initial speed given by the escape speed as determined in Section \ref{sec:num:find_momentum}.  The black hole enters the star supersonically, launching a shock wave and leaving strong density contrasts in its wake (top right).  For this particular collision the black hole is sufficiently large to accrete the entire star during its initial passage (bottom left), resulting in a black-hole remnant (bottom right).  In each panel we center the $z$-axis on the center of mass as determined in Section \ref{sec:num:com}, so that the origin $z = 0$, which is attached to the location of the black hole, appears at different locations in the different panels.  (See \cite{animation} for animations.)  }
    \label{fig:Run_B1}
\end{figure*}

\begin{figure*}
    \centering
    \includegraphics[width = 0.48 \textwidth]{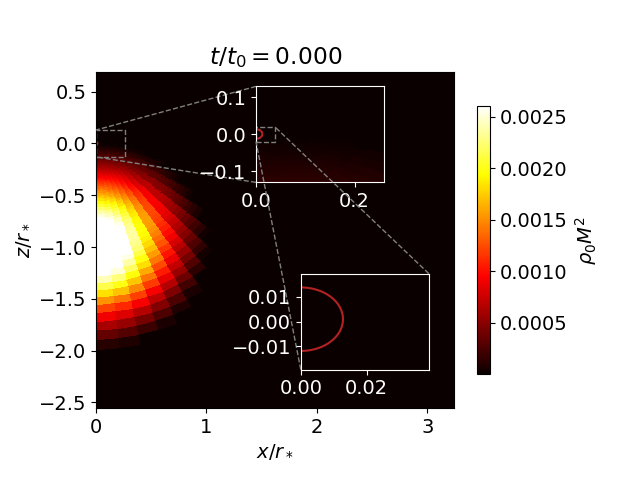}
    \includegraphics[width = 0.48 \textwidth]{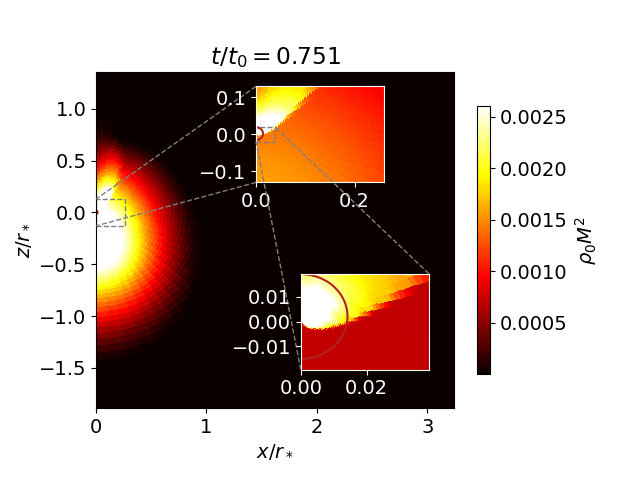}

    \includegraphics[width = 0.48 \textwidth]{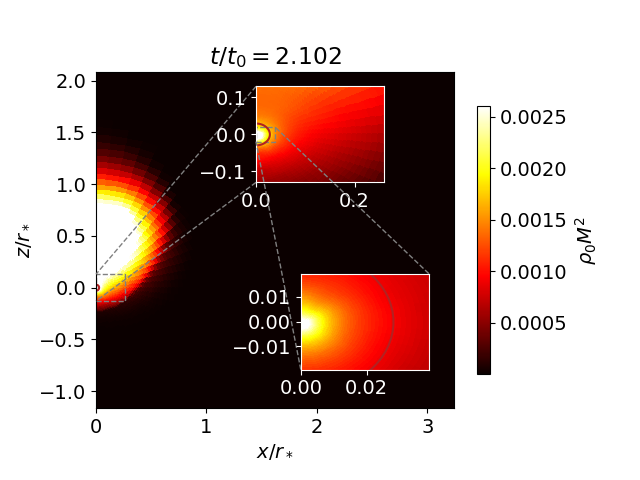}
    \includegraphics[width = 0.48 \textwidth]{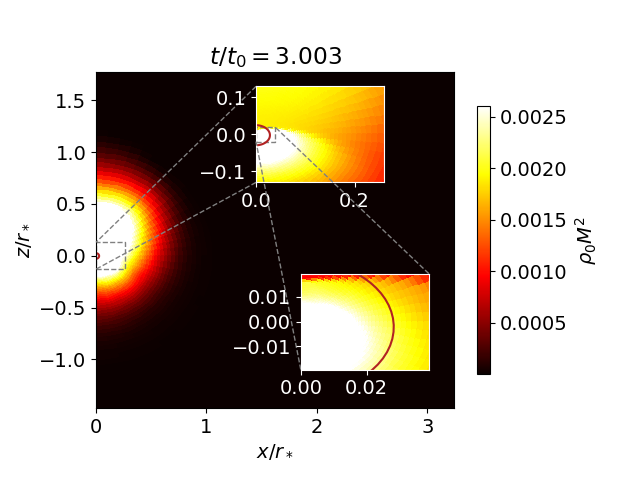}

    \includegraphics[width = 0.48 \textwidth]{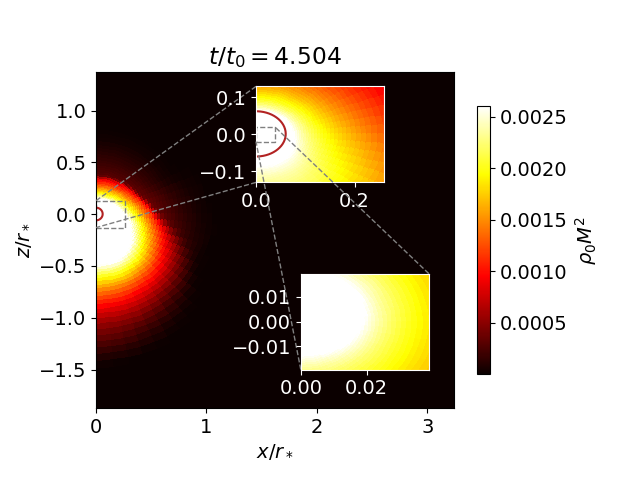}
    \includegraphics[width = 0.48 \textwidth]{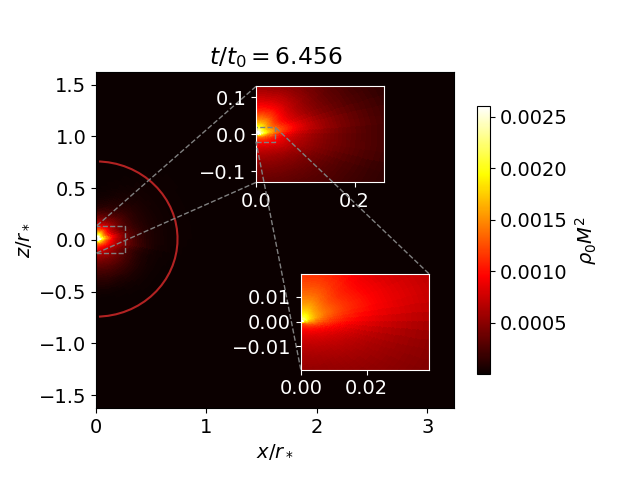}
    \caption{Same as Fig.~\ref{fig:Run_B1}, except for a black hole with initial mass $\bar m = 5 \times 10^{-3}$ (Case B2 in Table \ref{tab:transit}).  Here the black hole {\em almost} reemerges from the star around time $t \simeq 2 t_0$ after the first passage, but then turns around, nearly completes a second passage, and ends up accreting the entire star.  (See \cite{animation} for animations.) }
    \label{fig:Run_B2}
\end{figure*}

\begin{figure*}
    \centering
    \includegraphics[width = 0.48 \textwidth]{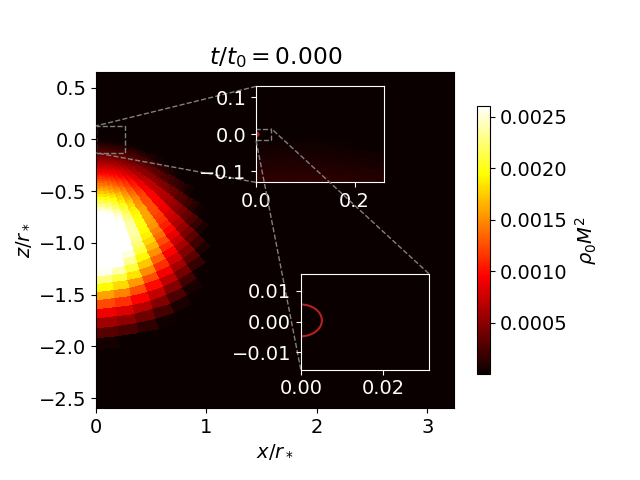}
    \includegraphics[width = 0.48 \textwidth]{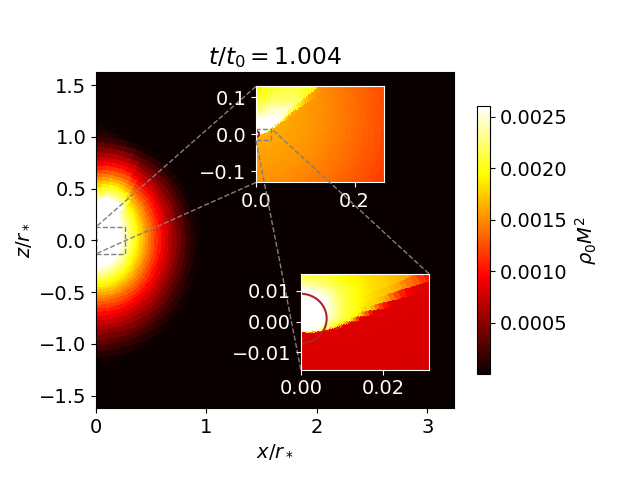}

    \includegraphics[width = 0.48 \textwidth]{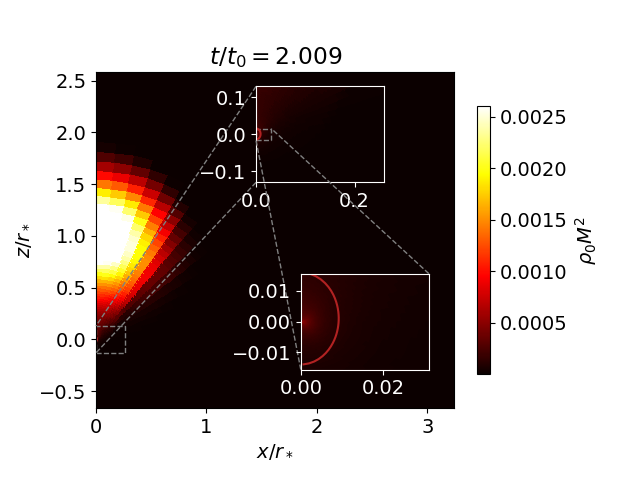}
    \includegraphics[width = 0.48 \textwidth]{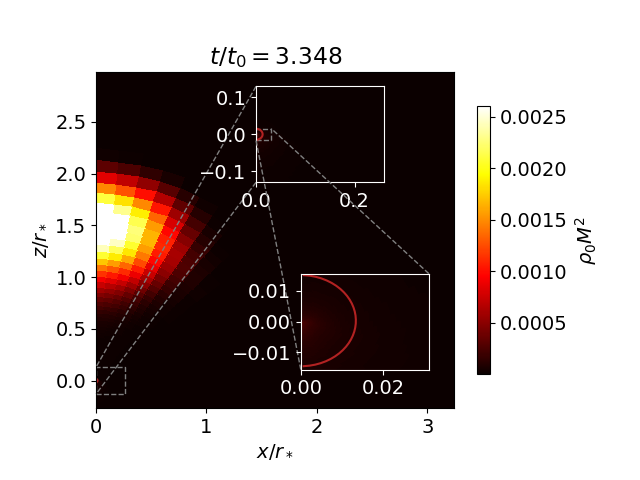}
    \caption{Same as Figs.~\ref{fig:Run_B1} and \ref{fig:Run_B2}, except for a black hole with initial mass $\bar m = 2 \times 10^{-3}$ (Case B3 in Table \ref{tab:transit}).  The black emerges from the star around $t \simeq 2 t_0$ without inducing a collapse of the neutron star.  (See \cite{animation} for animations.) }
    \label{fig:Run_B3}
\end{figure*}

In order to model the initial passage of a black hole through a neutron star we construct initial data describing the black hole at the stellar surface, entering the star with the escape speed as determined in Section \ref{sec:num:find_momentum}.   Whether or not this first transit leads to a collapse of the neutron star depends on whether the black hole had enough time to accrete at least a significant fraction of the stellar material.

Invoking the Newtonian estimates of Section II in Paper I we find the transit timescale $\tau_{\rm trans}$ from
\begin{equation} \label{tau_trans}
\tau_{\rm trans} \simeq \frac{2 R_*}{v_{\rm esc}} = 
\left( \frac{2 R_*^3}{M_*} \right)^{1/2} \simeq \sqrt{2} \, t_0,
\end{equation}
({\em cf.}~Eq.~4 in Paper I, hereafter I.4, where we neglected factors of order unity).  In order to estimate the accretion time scale $\tau_{\rm acc}$ we start with the relativistic expression for steady-state, spherical Bondi accretion,\footnote{Below we include a term in order to correct for the finite speed between the black hole and fluid (see Eq.~\ref{mdot_general}), and model the difference between rest-mass and mass-energy accretion (see Eq.~\ref{m_dot}; see also Appendix~\ref{sec:accretion}).}
\begin{equation} \label{Bondi}
\dot m_0^{\rm sph} = \frac{4 \pi \lambda_{\rm GR} \rho_0 m^2}{a^3},
\end{equation}
where $a$ is the speed of sound and $\lambda_{\rm GR}$ an accretion eigenvalue that, in general, depends on both $\Gamma$ and $a$ (see \cite{Bon52,Mic72}, as well as a textbook treatment in \cite{ShaT83}, for
$1 \leq \Gamma \leq 5/3$; see \cite{RicBS21a} and references therein for accretion of stiff EOSs with $\Gamma > 5/3$).  Both $\rho_0$ and $a$ are evaluated at a large distance from the black hole, where the fluid is assumed to be at rest.  Estimating $\tau_{\rm acc} \simeq m / \dot m_0$ and equating the result with $\tau_{\rm trans}$ then yields, up to factors of order unity, Eq.~(I.17), which can be evaluated to yield an approximate stability limit of $m \simeq 0.2 M_*$.

In order to explore transits close to the stability limit we consider collisions between black holes with masses $\bar m = 10^{-2}$, $5 \times 10^{-3}$, $2 \times 10^{-3}$, and $10^{-3}$, and the three neutron star models of Table \ref{tab:TOV_solutions}.  The physical parameters describing our twelve different initial data configurations, together with a characterization of the collision's outcome, are listed in Table \ref{tab:transit}.  We label each initial data set with a letter corresponding to the neutron star model, as in Table \ref{tab:TOV_solutions}, and a number labeling the black hole mass.

In Figs.~\ref{fig:Run_B1}, \ref{fig:Run_B2}, and \ref{fig:Run_B3} we show snapshots of density profiles for the Cases B1, B2, and B3, respectively. All three evolutions start with initial data, shown in the top left panels, describing the black holes at the surface of neutron star model B with  escape speed as determined in Section \ref{sec:num:find_momentum}.  The black holes enter the stars supersonically, launching shock waves that leave strong density contrasts in their wake.  As expected, the subsequent evolution depends strongly on the black hole mass.

For the largest black hole mass, case B1 with $m/M_0 = 0.0625$ shown in Fig.~\ref{fig:Run_B1}, the black hole accretes the stellar material very quickly, consuming much of the star and triggering stellar collapse before reaching the surface on the opposite side of the star.  The bottom right panel shows the remnant black hole, whose mass is very close to the total ADM mass of the spacetime.

In Fig.~\ref{fig:Run_B2} we show case B2 with a smaller black hole mass of $m / M_0 = 0.0313$.  Here the black hole {\em almost} reemerges on the opposite of the star, but then turns around and triggers dynamical collapse of the star close to the original surface of the star.  The bottom right panel shows the last timestep in our evolution, with almost all the mass absorbed by the black hole.

Finally, in Fig.~\ref{fig:Run_B3}, we show case B3 with a black hole mass of $m / M_0 = 0.0125$.  While the black hole still launches a strong shock wave and also accretes some stellar material, it reemerges from the star after a time of approximately $t \simeq 2 t_0$ without triggering an instability or collapse.  The bottom right panel shows the black hole and neutron star at time $t \simeq 3.35 t_0$, at which time the black hole has moved approximately an initial stellar radius away from the star, and the star is relaxing into a new equilibrium configuration, even though it is increasingly poorly resolved by our spherical polar coordinate system centered in the black hole. 

\begin{figure}
    \centering
    \includegraphics[width = 0.45 \textwidth]{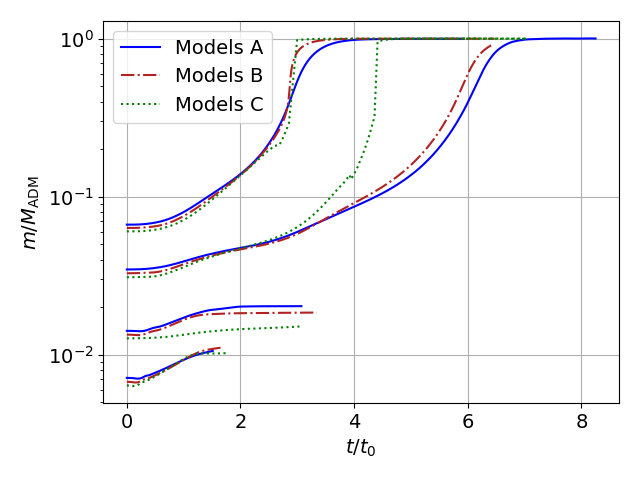}
    \caption{Black hole masses $m$ (in units of the total ADM mass $M$) as a function of time for the cases listed in Table \ref{tab:transit}.}
    \label{fig:mass_plot}
\end{figure}

In order to compare the different cases listed in Table~\ref{tab:transit} more systematically we show in Fig.~\ref{fig:mass_plot} the black hole masses $m$ (in units of the spacetimes' total ADM mass $M$) as a function of time.  Because of limitations related to the Courant time step, we evolved configurations with larger black hole masses to later times than those with smaller masses (which required a higher grid resolution close to the center).  While the results shown in Fig.~\ref{fig:mass_plot} are sure to be affected quantitatively by our coarse grid resolutions, we found that the qualitatively features do not depend on specific grid choices.  

In particular we find that, for black hole masses with $m/M_0 \simeq 0.06$  (Cases A1, B1, and C1), the neutron star collapses promptly and is completely accreted by the black hole during its first passage through the star, independently of how close the neutron star mass is to the maximum allowed mass.  In all three cases the black hole mass ends up very close to the spacetime's initial ADM mass, indicating that the entire neutron star has been absorbed by the black hole.   During the early part of the accretion process the black hole's growth rate depends only weakly on the neutron star mass and density, which is consistent with the properties of relativistic Bondi accretion for stiff equations of state (see Fig.~4 in \cite{RicBS21a}).  During the late, dynamical, phase of the accretion process, however, the black hole appears to grow more rapidly for larger neutron star masses.  

For black holes with masses $m/M_0 \simeq 0.03$ the black hole almost completes an entire passage through the star without triggering dynamical collapse, but then turns around and induces collapse during a subsequent passage.  In Table \ref{tab:transit} we mark this outcome as ``collapse after first passage".   

Finally, for smaller black hole masses (Cases A3, A4, B3, B4, C3, and C4) we find that the black hole reemerges from the neutron star without inducing collapse.  For all black hole masses considered here we expect the black hole to lose sufficient energy in this first passage to remain gravitationally bound to the neutron star, so that it would ultimately return and induce collapse during a subsequent passage.

From our numerical results we conclude that a black hole can indeed pass through a neutron star without inducing the ``pierced" star to collapse.  Moreover we observe that the crude Newtonian criterion (I.17) for collapse during the first passage, which we invoked at the beginning of this section, slightly overestimates the critical mass ratio $m / M_*$ below which the star remains stable.

%
\subsection{Late evolution}
\label{sec:res:late}
%

\begin{figure}
    \centering
    \includegraphics[width = 0.45 \textwidth]{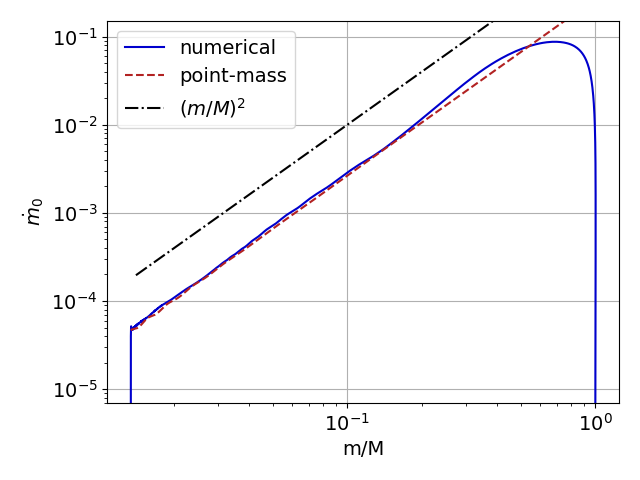}
    \caption{The (rest-mass) accretion rate $\dot m_0$ versus the black hole mass $m$ (in units of the spacetime's initial ADM mass $M$) for the late evolution simulation of Section \ref{sec:res:late}.  The solid (blue) line shows results from the numerical evolution of Section \ref{sec:res:late}, which evaluates $\dot m_0$ from (\ref{M_dot_3}), while the dashed (red) line is computed from the point-mass treatment of Paper I (see Eq.~I.43).  The dash-dotted (black) line shows the expected scaling $\dot m_0 \propto m^2$ based on the Bondi expression (\ref{Bondi}).} 
    \label{fig:accretion}
\end{figure}

We now turn to the late evolution of a PBH inside a neutron star, focusing on the last few dynamical timescales before collapse is induced.  

As shown in Fig.~11 of Paper I, a PBH with small initial mass sinks to small radii close to the center of the neutron star before its mass is large enough to trigger dynamical collapse.   The early phase of such an evolution, while $m \ll M_*$, can then be modeled with a point-mass treatment as in Paper I, while the late phase, including the final collapse, is best described by numerical simulations that place the black hole at the stellar center (see \cite{EasL19,RicBS21b,SchBS21}).  

Instead, we here consider black holes with masses large enough to almost trigger collapse, but that have not yet sunk to the stellar center.  As a specific example we construct initial data describing a neutron star with mass $\bar M_0 = 0.16$ (Model B of Table~\ref{tab:TOV_solutions}) harboring a black hole of mass $\bar m = 0.002$, initially at rest, at a radius $r_{\rm init} = 0.13 \, r_*$.  

In Fig.~\ref{fig:accretion} we show the (rest-mass) accretion rate $\dot m_0$ versus the black hole's mass $m$.  The numerical simulation, shown as the solid (blue) line, starts with zero accretion rate for the initial mass $m / M_* = 0.0125$, but quickly settles down to a steady-state accretion rate.  We also include results from the point-mass treatment of Paper I as the dashed (red) line.  This accretion rate is based on the Bondi rate (\ref{Bondi}), which is derived under the assumption that asymptotically the fluid is at rest with respect to the black hole.  In order to allow for a relative (asymptotic) speed $v$ between the two we include a Bondi-Hoyle-Lyttleton-like correction factor
\begin{equation} \label{mdot_general}
\dot m_0 = \dot m_0^{\rm sph} \left( \frac{a^2}{a^2 + v^2} \right)^{q/2} 
\end{equation}
(see, e.g., \cite{Bon52,ShiMTS85,PetSST89}).  As discussed in Paper I we adopt $q = 2$ for accretion with $\Gamma = 2$.  We note, however, that for the late evolution considered here we have $v/a \lesssim 0.3$, so that the effect of the correction factor relatively is small.  As demonstrated in Fig.~\ref{fig:accretion}, the two approaches agree well for much of the evolution, following the quadratic dependence of $\dot m_0$ on $m$ as indicated by the black dash-dotted line, until $m \simeq M_*$.  While the point-mas approach ignores any changes in the star, the accretion in the self-consistent simulation ceases, as expected, once all the mass has been accreted and the black hole's mass has reached the spacetime's initial ADM mass.

\begin{figure}
    \centering
    \includegraphics[width = 0.45 \textwidth]{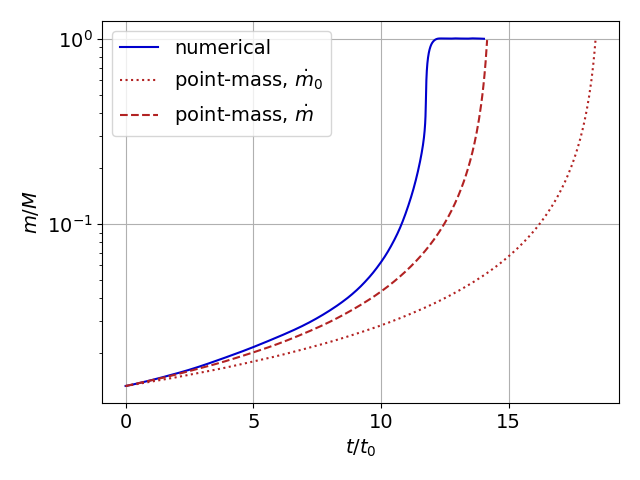}
    \caption{The black hole mass $m$ (in units of the ADM mass $M$ (as a function of time.  As in Fig.~\ref{fig:accretion}, the solid (blue) line shows results for the numerical simulation of Section \ref{sec:res:late}, while the dotted (red) line is based on the point-mass treatment of Paper I with the black hole growth approximated by $\dot m_0$ (see text for details).   The dashed (red) line is also based on the same point-mass treatment, but models the black hole growth from $\dot m$ using Eq.~(\ref{m_dot}).}
    \label{fig:m_versus_t}
\end{figure}

Next we show in Fig.~\ref{fig:m_versus_t} the black  hole mass $m$ as a function of time.  As expected, $m$ increases rather slowly initially, but then more rapidly as $m$ itself increases.   In the numerical simulations $m$ reaches a plateau once the black hole has accreted the entire star and its mass equals the initial ADM mass $M$, while in the point-mass treatment with the neutron star fixed we simply terminated the calculation.  While the qualitative behavior up to this point is very similar between the numerical and point-mass treatment of Paper I, shown as the dotted (red) line in Fig.~\ref{fig:m_versus_t}, we observe that the latter appears to proceed noticeably more slowly than the former (compare also Fig.~8 in \cite{RicBS21b}).  

At least in part, this is difference due to the fact that the point-mass treatment of Paper I models the growth of the black hole from the accretion rate (\ref{mdot_general}), and hence accounts for the accretion of rest-mass only.  In reality, the black hole mass $m$ grows in response to the accretion of all forms of mass-energy, and generally will therefore grow faster than predicted by (\ref{mdot_general}).  Since $m$ appears on the right-hand side of the Bondi-accretion rate (\ref{Bondi}), an underestimate of $m$ will reduce the accretion rate, and therefore prolong the evolution.  Following \cite{AguST21,AguTSL21}, whose arguments we review in Appendix \ref{sec:accretion}, the rate of mass-energy accretion can be better estimated by comparing the rest-mass current with the mass-energy current, resulting in the simple expression
\begin{equation} \label{m_dot}
    \dot m \simeq h \dot m_0
\end{equation}
(see \ref{m_dot_app}).   Here the specific enthalpy
\begin{equation} \label{enthalpy}
h \equiv \frac{\rho + P}{\rho_0} = 1 + \frac{\Gamma}{\Gamma - 1} \, \frac{P}{\rho_0}
\end{equation}
is evaluated at a large distance from the black hole, and the second equality holds for a $\Gamma$-law EOS (\ref{Gamma_law}).  Using $\dot m$ based on (\ref{m_dot}), rather than $\dot m_0$ based on (\ref{mdot_general}), results in the dash-dotted (red) line in Fig.~\ref{fig:m_versus_t}, labeled as $\dot m$.  Evidently, accounting for the difference between rest-mass and mass-energy accretion in the point-mass treatment results in significantly better agreement with our numerical results. We repeat that our full numerical simulation of the black hole--neutron star spacetime adopts no approximation in determining the black hole growth.

\begin{figure}
    \centering
    \includegraphics[width = 0.45 \textwidth]{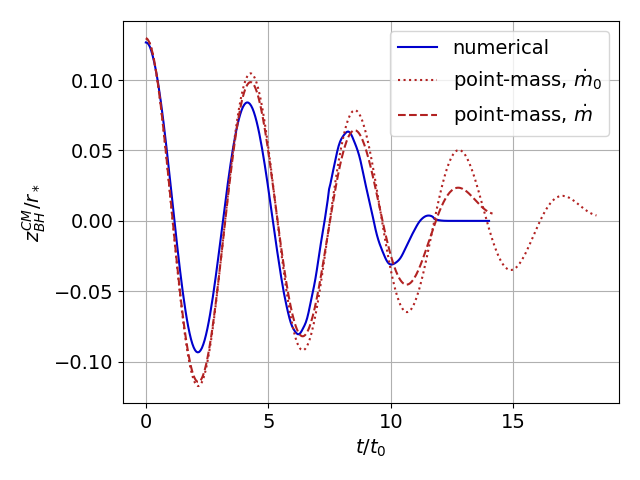}
    \caption{The coordinate location of the black hole (relative to the center of mass) as function of time for the late evolution of Section \ref{sec:res:late}.  As in Figs.~\ref{fig:accretion} and \ref{fig:m_versus_t}, the solid (blue) line shows numerical results while the dotted and dashed (red) lines show results from the point-mass treatment of the Paper I.  For the former the  center of mass is defined as in (\ref{com}), while, for the latter, the center of mass is well approximated by the center of the star.}
    \label{fig:trajectory}
\end{figure}

Finally, we show in Fig.~\ref{fig:trajectory} the black hole's coordinate location $z^{\rm CM}_{\rm BH}$ as a function of time.  The solid (blue) line again denotes numerical simulation results, while the dotted and dashed (red) lines show results from the test-mass treatment, using either $\dot m_0$ (dotted) or $\dot m$  (dashed) to model black hole growth.  While, not surprisingly, the different approaches result in some quantitative differences, they all display the expected behavior resembling a damped oscillation.  Invoking a simple Newtonian argument we can estimate the period of this oscillation as follows: The restoring force acting on the black hole at a distance $r$ from the center is, to leading order, that of a harmonic oscillator, $F = m M(r) / r^2 \simeq 4 \pi m \rho_c r / 3 = k r$, where $M(r)$ is the enclosed mass, we defined
\begin{equation}
k \equiv \frac{4 \pi m \rho_c}{3} = 
\frac{4 \pi m \delta\rho_{\rm ave}}{3} = 
m \delta \frac{M}{R^3} = \frac{m \delta}{t_0^2},
\end{equation}
and $\delta = \rho_c / \rho_{\rm ave}$ is the central condensation.  Accordingly, the period of the oscillation is given by
\begin{equation}
P = \frac{2 \pi}{\omega} = 2 \pi \left(\frac{m}{k} \right)^{1/2} = \frac{2 \pi}{\sqrt{\delta}} t_0.
\end{equation}
With $\delta = 3.29$ for a $\Gamma = 2$ Newtonian polytrope we have 
$P \simeq 3.5 t_0$, in good agreement with the results shown in Fig.~\ref{fig:trajectory}.  Moreover, consistent with Fig.~\ref{fig:m_versus_t}, we observe that the numerical and point-mass results for the rate of damping agree better if the latter account for mass-energy accretion (\ref{m_dot}) rather than just rest-mass accretion (\ref{mdot_general}) in the black-hole growth.

%
\section{Summary}
\label{sec:discussion}
%

We present axisymmetric dynamical simulations of (potentially primordial) black holes inside neutron stars, self-consistently solving Einstein's equations together with the equations of relativistic hydrodynamics.  Our approach here complements that of Paper I, where we assumed that the black hole mass $m$ is much smaller than the mass $M_*$ of the host star.  In that case the effects of the black hole on the neutron star can be neglected and the orbit and evolution of the black hole can be modeled within a point-mass treatment.  Here, on the other hand, we allow for larger black-hole masses, so that the effects of the black hole on the neutron star cannot be neglected, and instead have to be modeled in a self-consistent simulation of both black hole and neutron star.

We construct initial data describing boosted black holes either inside or outside neutron stars, and adopt a relativistic prescription for determining the black hole's boost corresponding to the escape speed.  We also describe a ``fix-point" shift condition that keeps the black hole at a fixed coordinate location throughout the evolution, allowing us to take advantage of the high radial resolution close to the origin in our spherical polar coordinate system.

We then consider two different types of scenarios.  In a first set of simulations we model the head-on collision of black holes with neutron stars.  We start these simulations with data describing the black hole at the stellar surface, about to enter the star with escape speed.  In particular we demonstrate that, for sufficiently small black-hole masses $m$, the black hole passes through the star without triggering collapse, meaning that a neutron star can indeed remain stable despite having been pierced by a black hole.  However the black hole remains bound to the neutron star and will return to penetrate it again.  For larger black hole masses $m$ the neutron star collapses after one more more passages, in approximate agreement with Newtonian estimates.  The transition occurs at about $m \approx 0.05 M_*$, roughly independent of the neutron star mass $M_*$.

As a second scenario we consider the late evolution, with the black hole oscillating about the stellar center just before triggering dynamical collapse.  We track the oscillation for about three periods until the accretion becomes dynamical and results in the entire star being quickly swallowed by the black hole.  During the early part of these evolutions our results agree well with those from the point-mass treatment, especially if the latter model the black-hole growth based on the accretion of mass-energy rather than rest-mass.


\acknowledgments

T.W.B.~gratefully acknowledges hospitality at the University of Illinois at Urbana-Champaign's Center for Advanced Studies of the Universe.  This work was supported in part by National Science Foundation (NSF) grants PHY-2010394 and PHY-2341984 to Bowdoin College, as well as NSF grants PHY-2006066 and PHY-2308242 to the University of Illinois at Urbana-Champaign.

\begin{appendix}
%
\section{Rest-mass versus mass-energy in Bondi accretion}
\label{sec:accretion}
%

The accretion rate (\ref{Bondi}) for adiabatic spherical flow in steady-state for a perfect gas at rest and homogeneous at infinity (i.e.~relativistic Bondi accretion) follows from the continuity equation and the existence of a time-like Killing vector $\xi^a_{(t)}$.  The continuity equation, i.e.~the conservation of the rest-mass four-current $J^a = \rho_0 u^a$, gives 
\begin{equation} \label{divJ}
 \nabla_a J^a = \nabla_a (\rho_0 u^a) = 0.
\end{equation}
Evaluating Eq.~\ref{divJ} for spherical flow in steady-state yields
\begin{equation}
    \partial_r ( \sqrt{-g} \rho_0 u^r ) = 0,
\end{equation}
where $g$ is the determinant of the spacetime metric. Neglecting the self-gravity of the fluid and adopting Schwarzschild coordinates for the central black hole gives $g = - r^2 \sin\theta$, from which we conclude that 
\begin{equation}
    \rho_0 u^r r^2 = \mbox{const}.
\end{equation}
The rest-mass accretion rate through a surface of radius $r$,
\begin{equation} \label{restmassflux}
    F = \int J^r r^2 d \Omega = 
    4 \pi \rho_0 u^r r^2,
\end{equation}
is therefore independent of $r$ and thus applies at the horizon.  In particular we conclude that the rest-mass $m_0$ enclosed within $r$ increases at a rate given by  (\ref{restmassflux}),
\begin{equation} \label{M0_dot}
\dot m_0 = \frac{d m_0}{dt} = F = 4 \pi \rho_0 u^r r^2.
\end{equation}
Evaluating the fluid equations for stationary flow admits the existence of a critical point (where, in the Newtonian limit, the fluid becomes supersonic).  This critical solution must apply for any equation of state (EOS) obeying the causality constraint, $a^2 < 1$, where $a$ is the sound speed~\cite{ShaT83}.  Imposing regularity at this critical point determines the flux (\ref{restmassflux}) and yields the  accretion rate (\ref{Bondi}) (see \cite{Bon52,Mic72}, as well as \cite{ShaT83} for a textbook treatment, for $1 \leq \Gamma \leq 5/3$; see \cite{RicBS21b} for accretion of stiff EOSs with $\Gamma > 5/3$).  By construction, the Bondi accretion rate yields the rate of {\em rest-mass} accretion.

However, a black hole's mass $m$ grows in response to the accretion of {\em all forms of mass-energy}.  While the rate of this growth can be evaluated rigorously in the context of the dynamical-horizon formalism (see, e.g.,~\cite{AshK03}), we here invoke a simpler, approximate argument to estimate this rate (see also \cite{AguST21,AguTSL21}).  Specifically, we again assume the existence of an (approximate) time-like Killing vector $\xi^a_{(t)}$, in which case the energy four-current ${\mathcal J}^a \equiv T^{a}_{~b} \xi^b_{(t)} = T^a_{~t}$ is approximately conserved,
\begin{equation} \label{divcurlyJ}
     \nabla_a {\mathcal J}^a = \nabla_a \left((\rho + P) u^a u_t + P g^a_{~t} \right) = 0.
\end{equation}
Following the same arguments as above we obtain, for a perfect gas,
\begin{equation}
    \partial_r \left( (\rho + P) u^r u_t r^2 \right) = 0,
\end{equation}
so that the mass-energy accretion rate through a surface of radius $r$
\begin{equation}
    {\mathcal F} = \int {\mathcal J}^r r^2 d \Omega =  4 \pi (\rho + P) u^r u_t r^2
\end{equation}
is again independent of $r$, and gives rise to a change in the enclosed energy $\mathcal E$ given by 
\begin{equation} \label{E_dot_1}
    \dot {\mathcal E} = {\mathcal F} = 4 \pi (\rho + P) u^r u_t r^2.
\end{equation}

We can simplify the right-hand side of (\ref{E_dot_1}) by first using the normalization of the four-velocity $u^a$, which yields 
\begin{equation} \label{u_norm}
    u_t = \left(1 - \frac{2M}{r} + (u^r)^2 \right)^{1/2}
\end{equation}
in Schwarzschild coordinates.  We next use the Bernoulli equation
\begin{align} \label{Bernoulli}
    & \frac{\rho + p}{\rho_0} \left(1 - \frac{2M}{r} + (u^r)^2 \right)^{1/2} = \mbox{const} =
    \frac{\rho_{\infty} + p_{\infty}}{\rho_{0\infty}} \nonumber \\
    & ~~~~~ = 
    1 + \frac{\Gamma}{\Gamma - 1} \frac{P_{\infty}}{\rho_{0\infty}}
    = h_{\infty}
\end{align}
(see, e.g., Eq.~G.22 in \cite{ShaT83}).  Here the subscript $\infty$ refers to $r \rightarrow \infty$,
we have assumed $u^r \rightarrow 0$ in this limit, have adopted the Gamma-law EOS (\ref{Gamma_law}), and in the last equality we introduced the specific enthalpy $h \equiv (\rho + P) / \rho_0$.  Using both (\ref{u_norm}) and (\ref{Bernoulli}) in (\ref{E_dot_1}) we now obtain
\begin{equation} \label{E_dot_2}
    \dot {\mathcal E} = 4 \pi \rho_0 h_\infty u^r r^2 = h_\infty \dot m_0
\end{equation}
(compare Eq.~42 in \cite{AguST21}, where, by their Eq.~7, their variable $e$ equals $h_\infty$ when $u^r_\infty = 0$).  We finally approximate the rate of black-hole growth, $\dot m$, as $\dot {\mathcal E}$ to obtain the result
\begin{equation} \label{m_dot_app}
\dot m \simeq h_\infty \dot m_0.
\end{equation}
We note that ${\dot m} \geq {\dot m}_0$ as expected.   We also observe that we can have $\dot m > 0$ even if the rest-mass density $\rho_0$ and hence $\dot m_0$ both vanish while the pressure remains non-zero -- as, for example, for a photon gas.

As a numerical example we refer to Table II of \cite{RicBS21b}, which reports numerical simulations of the accretion onto a black hole at the center of the neutron star.  Adopting their values at the center of the neutron star, namely $\bar \rho_{0c} = 0.2$ and $\Gamma = 2$, as the (local) asymptotic boundary conditions for relativistic Bondi flow,  we have $h_\infty = 1.4$.  Computing ratios between $\dot m$ and $\dot m_0$ from the values provided in Table II, we find $\dot m / \dot m_0 \simeq 1.35$.  We conclude that the estimate (\ref{m_dot}) based on mass-energy flow indeed provides a significantly better approximation to the increase in the black-hole mass $m$ than the relativistic Bondi expression (\ref{M0_dot}), which is based on rest-mass flow.

\end{appendix}


%

\end{document}